\documentclass[showpacs,amsmath,nofootinbib,amssymb,twocolumn,superscriptaddress]{revtex4-1}
\usepackage{array}
\usepackage{color}
\usepackage{dcolumn}
\usepackage{bm}
\usepackage{epsf}
\usepackage{graphicx}
\usepackage{amsmath} 

\newcommand{\beq}{\begin{equation}}
\newcommand{\eeq}{\end{equation}}
\newcommand{\bean}{\begin{eqnarray*}}
\newcommand{\eean}{\end{eqnarray*}\noindent}
\newcommand{\bea}{\begin{eqnarray}}
\newcommand{\eea}{\end{eqnarray}\noindent}

\begin{document}
\topmargin0in
\textheight 8.in 
\bibliographystyle{apsrev}
\title{Recent advances in neutrino astrophysics}

\author{Cristina Volpe}
\email{volpe@apc.univ-paris7.fr}
\affiliation{AstroParticule et Cosmologie (APC), Universit\'e Paris Diderot - Paris 7, 10, rue Alice Domon et L\'eonie Duquet, 75205 Paris cedex 13, France}

\begin{abstract}

Neutrinos are produced by a variety of sources that comprise our Sun, explosive environments such as core-collapse supernovae, the Earth and the Early Universe. The precise origin of the recently discovered ultra-high energy neutrinos is to be determined yet. These weakly interacting particles give us information on their sources, although the neutrino fluxes can be modified when neutrinos traverse an astrophysical environment. Here we highlight  recent advances in neutrino astrophysics and emphasise the important progress in our understanding of neutrino flavour conversion in media.
\end{abstract}

\date{\today}


\maketitle

\noindent
Neutrinos are intriguing weakly interacting particles that can travel over long distances to tell us properties of the environments that produce them. These elusive particles have kept misterious for a long time. After 1998 many unknown $\nu$ properties have been determined thanks to the discovery of neutrino oscillations, first proposed in \cite{Pontecorvo:1957cp}, an observation by the Super-Kamiokande experiment using atmospheric neutrinos \cite{Fukuda:1998mi}. This discovery is fundamental for  particle physics, for astrophysics and cosmology.

Neutrino oscillations is an interference phenomenon among the $\nu$ mass eigenstates, that occurs if neutrinos are massive and if the matter (propagation) basis and the flavor (interaction) basis do not coincide. The matrix that relates the two basis is called the Maki-Nakagawa-Sakata-Pontecorvo matrix \cite{Maki:1962mu}
which depends on three mixing angles, one Dirac  and two Majorana CP violating phases (if only three active neutrinos are considered).
Solar, reactor and accelerator experiments have determined  $\Delta m_{23}^2 = m_{3}^2 -  m_{2}^2 = 7.6 \times 10^{-3} $eV$^2$,  and $\Delta m_{12}^2 = m_{2}^2 -  m_{1}^2 = 2.4 \times 10^{-5} $eV$^2$, referred to as  the atmospheric and the solar mass-squared differences respectively.
Moreover the sign of $\Delta m^2_{12}$ has been measured by the occurrence of the Mikheev-Smirnov-Wolfenstein (MSW) effect in the Sun \cite{Wolfenstein:1977ue,Mikheev:1986gs}.
The sign of $\Delta m_{23}^2$ is still unknown so that there two are possible ways to order the mass eigenstates. 
The lightest mass eigenstate is $m_1$, if $\Delta m_{31}^2 > 0$   (normal ordering or "hierarchy"), or $m_3$ if  $\Delta m_{31}^2 < 0$ (inverted ordering).
Most of neutrino oscillation experiments can be interpreted within the framework of three active neutrinos. However a few measurements present anomalies that require further clarification. While no hypothesis can explain all anomalies, a seriously investigated explanation is provided by the existence of extra sterile neutrinos, that would not couple to the gauge bosons. Such neutrinos would then manifest themselves in oscillation experiments through the mixing with the other active species.  

Among the fundamental properties yet to be determined are the mechanism for the neutrino mass, the absolute mass value and ordering, the neutrino nature (Dirac versus Majorana), the existence of CP violation in the lepton sector and of sterile neutrinos. In the coming decade(s) experiments will determine some of the still unknown neutrino properties. Cosmological and astrophysical neutrinos offer possible ways in this quest to learn about fundamental neutrino properties. Coming from various sources, their energy 
spans  from meV to PeV. Their investigation is also pursued to learn about the sources that produce them (see e.g. \cite{Volpe:2013kxa}).  
Here we will highlight some recent advances in the fascinating field of neutrino astrophysics.

\noindent
{\bf Solar neutrinos}\\
Electron neutrinos are constantly produced in our Sun through the proton-proton (pp) nuclear reaction chain that produces 99 $\%$ of the Sun's energy by burning hydrogen into helium-4  \cite{Bethe:1939bt}. The corresponding solar neutrino flux receives contributions from both fusion reactions and beta-decays of $^{7}$Be and $^{8}$B. First measured by R. Davis pioneering experiment \cite{Davis:1968cp}, the solar $\nu$ flux was found to be nearly a factor of three below predictions \cite{Bahcall:1968hc}. Over the decades solar neutrino experiments  have precisely measured electron neutrinos from the different pp branches, usually known as the pp, pep, $^{7}$Be and $^{8}$B and hep neutrinos. 
The measurement of a "solar neutrino deficit" (a reduced solar neutrino flux compared to standard solar model predictions) has been confirmed by solar experiments mainly sensitive to electron neutrinos, but with some sensitivity to the other flavors. 

The proposed solutions questioned the standard solar model and included unknown neutrino properties (e.g. flavor oscillations and the MSW effect, a neutrino magnetic moment, neutrino decay). The MSW effect is due to the neutrino interaction with matter and the presence of resonant flavor conversion when they traverse a medium. This  can occur depending on the environment density profile on one hand and
neutrino properties on the other (energy, mass-squared differences value and sign, mixing angles).  If the evolution at resonance is adiabatic, electron neutrinos can efficiently convert into muon and tau neutrinos engendering a flux deficit.

The solar puzzle is definitely solved by the discovery of the neutrino oscillations  and the SNO and KamLAND experiments (see \cite{Robertson:2012ib} for a review). In fact, using elastic scattering, charged- and neutral- current neutrino interactions on heavy water, SNO has measured the total $^{8}$B solar neutrino flux consistent with predictions : solar electron neutrinos convert into the other active flavours. In particular, the muon and tau neutrino components of the solar flux have been measured at 5 $\sigma$ \cite{Ahmad:2002jz}. Moreover KamLAND has definitely identified the Large Mixing Angle (LMA) solution, by observing reactor electron anti-neutrino disappearance at an average distance of 200 km \cite{Eguchi:2002dm}. The ensemble of these observations shows that low energy solar neutrinos are suppressed by averaged vacuum oscillations while neutrinos having more than 2 MeV energy 
are suppressed by the MSW effect (Figure 1). Theoretically one expects $P (\nu_e \rightarrow \nu_e)  \approx 1 - {1 \over 2} \sin^2 2 \theta_{12} \approx 0.57 $
(with $\theta_{12} = 34^{\circ}$) for ($< 2$ MeV) solar neutrinos. 
For the high energy portion of the $^{8}$B spectrum, the matter-dominated survival probability is $P (\nu_e \rightarrow \nu_e) ^{high~density}  \rightarrow \sin^2\theta_{12} \approx 0.31$ (see e.g.\cite{Robertson:2012ib}).
\begin{figure}
\begin{center}
\includegraphics[width=0.5\textwidth]{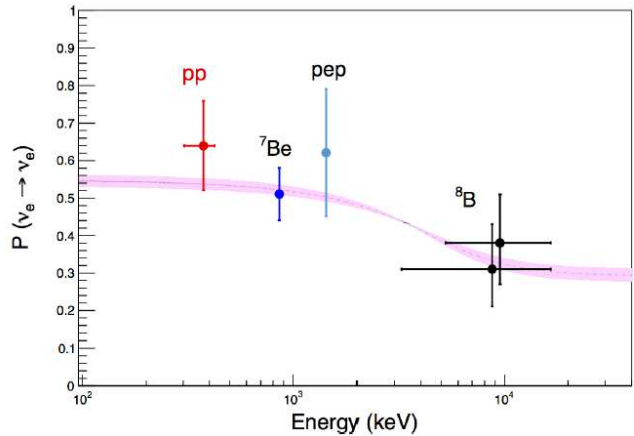}
\caption{Solar neutrinos : Electron neutrino survival probability, as a function of the neutrino energy, for the pp, pep, $^{7}$Be, $^{8}$B neutrinos from the Borexino experiment. The results are compared to averaged vacuum oscillation ($E_{\nu} < 2$ MeV) and the MSW ($E_{\nu} > 2$ MeV) predictions, taking into account present uncertainties on mixing angles. Figure from \cite{Bellini:2014uqa}. }
\label{fig1}
\end{center}
\end{figure}
Recently Borexino experiment has precisely measured the low energy part of the solar neutrino flux : the pep \cite{Collaboration:2011nga}, $^{7}$Be \cite{Arpesella:2008mt} and pp. In fact, by achieving challenging reduced backgrounds, the collaboration has just reported the first direct measurement
of pp neutrinos, the keystone of the fusion process in the Sun. The measured flux is consistent with the standard solar model predictions  \cite{Bellini:2014uqa}.

The ensemble of these observations has established that the Sun produces $3.84 \times 10^{33}$ ergs/s via the pp chain. Moreover the occurrence of the MSW effect for the high energy solar neutrinos shows that these particles change flavor in vacuum in a very different way than in matter. The MSW phenomenon occurs in numerous other contexts, including the early universe (at the epoch of the primordial abundances formation\footnote{in presence of sterile $\nu$}), massive stars like core-collapse supernovae,  accretion disks aroung black holes and
the Earth. Future measurements will aim at observing solar neutrinos produced in the Carbon-Nitrogen-Oxygen (CNO) cycle which is thought to be the main mechanism for energy production in massive stars  \cite{Bethe:1939bt}. Borexino has provided the strongest constraint on the CNO cycle, consistent with predictions \cite{Collaboration:2011nga}. Moreover the precise determination of the transition between the vacuum averaged and the LMA solution would give valuable information since deviations from the simplest vacuum-LMA transition could point to new physics, such as non-standard neutrino interactions \cite{Friedland:2004pp}.   

\noindent
{\bf Neutrinos from core-collapse supernovae}\\
Massive stars are powerful neutrino sources. At the end of their life they shine 99$\%$ of their gravitational energy (about 3$\times$10$^{53}$ ergs) as neutrinos of all flavors in a burst
that lasts 10 s. About twenty events have been observed for the first time from this type of sources during the explosion of the SN1987A located in the Large Magellanic Cloud at 10 kpc from the Earth (see  \cite{Suzuki:2008zzf}, \cite{Vissani:2014doa} for a review). Although limited in statistics these events have roughly confirmed  the predictions on the neutrino luminosity curve from a supernova and furnished interesting information on neutrino properties and fundamental interactions. For example, a constraint on CPT and Lorentz violation is obtained from the nearly simultaneous arrival of photons and neutrinos \cite{Longo:1987ub}.
 
The SuperNova Early Warning System (SNEWS) and numerous other neutrino detectors around the world can serve as supernova neutrino observatories, if a supernova blows up in the Milky Way or outside our galaxy. Large scale detectors based on different technologies \cite{Autiero:2007zj,Scholberg:2012id} including liquid argon, water Cherenkov and scintillators are being considered (e.g. JUNO or Hyper-K). These have the potential to detect neutrinos from a galactic and an extragalactic explosion as well as to discover the diffuse supernova neutrino background produced from explosions up to cosmological redshift of 2 (for a review see \cite{Beacom:2010kk,Lunardini:2010ab}). 

Supernova neutrinos are tightly connected with two major questions in astrophysics, namely what is the mechanism that makes massive stars explode and what is (are) the site(s) where the heavy elements are formed. Significant progress has been performed with the achievement of multi-dimensional simulations that include turbulence, convection, hydrodynamical instabilities like the standing-accretion-shock, realistic nuclear networks and neutrino transport (see \cite{Janka:2012wk} and \cite{Burrows:2012ew} for reviews). 
Currently simulations for several progenitors, from different groups, show indications for successful explosions in 2-dimensions.  
Calculations based on 3-dimensions are just appearing which has uncovered a new neutrino-hydrodynamical instability  termed LESA  (Lepton-number Emission Self-sustained Asymmetry)  \cite{Tamborra:2014aua}. In the neutrino driven mechanism, neutrinos significantly contribute to the explosion by depositing energy behind the shock.
Instabilities are also thought to play an important role. Identifying the exact interplay among different features, and if these are sufficient to obtain successful explosions
in three-dimensions, is an open issue.
   
Besides astrophysical conditions and exotic nuclei properties, neutrinos also contribute to determine heavy element abundances. Core-collapse supernovae, accretion-disks around black holes and neutron-star mergers are among the candidate sites for the $r$-process. The interaction of electron neutrinos and anti-neutrinos with neutrons and protons in such environments determines the neutron-to-proton ratio, a key r-process parameter. While numerous studies show that neutrinos impact the neutron richness of a given astrophysical environment, finally assessing their influence still requires extensive simulations that e.g. self-consistently include neutrino and matter evolution (for recent reviews see \cite{Volpe:2014yqa}). 
  
Interestingly the study of massive stars has revealed new facts compared to the case of the Sun.
In fact, besides the MSW effect \cite{Dighe:1999bi}, recent calculations have shown the emergence of new phenomena due to the neutrino-neutrino interaction, the presence of shock waves and of turbulence (see \cite{Duan:2010bg,Duan:2009cd} for a review). 
The role of the neutrino interaction with themselves, first pointed out in \cite{Pantaleone:1992eq}, has been shown to influence the r-process and to impact the neutrino fluxes  with sharp spectral changes. New flavour conversion regimes have been identified (synchronisation, bipolar oscillations and spectral split) and the occurrence of collective stable and unstable modes of the (anti-)neutrino gas in the star. The underlying mechanisms are often investigated using the evolution of effective (iso)spin in effective magnetic fields.  This has allowed for example to show that the spectral split phenomenon is an MSW-effect in a comoving frame \cite{Raffelt:2007xt}, or analogous to a magnetic-resonance phenomenon \cite{Galais:2011gh}.  On the other hand steep changes of the stellar density profile due to shock waves induce multiple MSW resonances and interference among the matter eigenstates, rendering neutrino evolution completely non-adiabatic. 

Important progress has been accomplished in our understanding of how neutrinos change their flavour in these sites and many general features established. Still further work is needed to address important questions, to assess e.g. what is their final impact on a supernova neutrino signal on Earth, or how these findings are modified in an improved theoretical framework based on wave-packet treatment \cite{Akhmedov:2014ssa} or beyond the mean-field approximation \cite{Balantekin:2006tg,Volpe:2013uxl,Vlasenko:2013fja} used so far. 
Extended descriptions describing neutrino evolution in a dense medium have recently been derived using either a coherent-state path integral \cite{Volpe:2013uxl}, or the Born-Bogoliubov-Green-Kirkwood-Yvon hierarchy \cite{Volpe:2013uxl}, or the two-particle-irreducible effective action formalism \cite{Vlasenko:2013fja} (see also \cite{Sigl:1992fn,McKellar:1992ja}). In an extended mean-field description
two kinds of corrections  have been identified : spin or helicity coherence \cite{Vlasenko:2013fja}  and neutrino-antineutrino pairing correlations \cite{Volpe:2013uxl}. The former are present because of the neutrino mass, the latter appear in an extended mean-field description with all possible two-point correlators. The most general equations for neutrino propagation including both corrections have been derived in \cite{Serreau:2014cfa}. While the origin is very different, both kinds of correlations introduce neutrino-antineutrino mixing, in presence of spatial anisotropies of the matter and/or neutrino backgrounds. Such corrections are expected to be tiny, but the non-linearity of the equations could introduce significant changes of neutrino evolution in particular in the transition region in supernovae. This is between the dense region (within the neutrinosphere) which is usually Boltzmann treated and the diluted one (outside the neutrinosphere) where collective flavor conversion occurs. So far, this transition has been treated as a sharp boundary where the neutrino fluxes and spectra obtained in supernova simulations are used as initial conditions for flavor studies.
Numerical calculations are needed to investigate the role of spin coherence, of neutrino-antineutrino pairing correlations, or of collisions.
A first calculation in a simplified model shows that helicity coherence might have an impact \cite{Vlasenko:2014bva}. 

Another interesting theoretical development is the establishment of connections between supernova neutrinos and many-body systems in other domains.
Using algebraic methods, Ref.\cite{Pehlivan:2011hp} has shown that the neutrino-neutrino interaction Hamiltonian can be rewritten as a (reduced) 
Bardeen-Cooper-Schrieffer (BCS) Hamiltonian for superconductivity  \cite{Bardeen:1957mv}. As mentioned above, Ref.\cite{Volpe:2013uxl} has included neutrino-antineutrino correlations of the pairing type, which are formally analogous to the BCS correlations. The linearisation of the corresponding neutrino evolution equations  has uncovered the formal link between stable or unstable collective neutrino modes and those well known in atomic nuclei and metallic clusters \cite{Vaananen:2013qja}.  

The observation of the neutrino luminosity curve from a future (extra)galactic explosion would closely follow the different phases of the explosion furnishing a crucial test of
supernova simulations and providing information on unknown neutrino properties. In particular, the occurrence of the MSW effect in the outer layers of the star
and of collective effects depends on the value of the third neutrino mixing angle and of the neutrino mass ordering. The precise measurement of the last mixing angle \cite{Abe:2011fz,An:2012eh,Ahn:2012nd} reduces the number of unknowns. Still, the neutrino signal from a future supernova explosion could tell us about the mass ordering, either from  the early time signal in ICECUBE \cite{Serpico:2011ir},  or  in Cherenkov, or scintillator detectors, by measuring the positron time and energy signal associated with the passage of the shock wave in the MSW region \cite{Gava:2009pj}. Several other properties can impact the neutrino fluxes such as the neutrino magnetic moment \cite{deGouvea:2012hg}, non-standard interactions, sterile neutrinos. CP violation effects from the Dirac phase exist but appear to be small \cite{Balantekin:2007es,Gava:2008rp,Kneller:2009vd,Pehlivan:2014zua}. In spite of the range of predictions, the combination of future observations using different detection channels that measure  flavour, time and energy with different thresholds can pin down degenerate solutions  (see e.g. \cite{Vaananen:2011bf}) and bring key information to this domain. 

\noindent
{\bf Ultra-high energy neutrinos}\\
The main mission of  high-energy neutrino telescopes is to search for galactic and extra-galactic sources of high-energy neutrinos to elucidate the source of cosmic rays and the astrophysical mechanisms that produce them. These telescopes also investigate neutrino oscillations, dark matter and supernova neutrinos (for IceCube). The 37 events collected in ICECUBE, with deposited energies ranging from 30 to 2 PeV,  is consistent with the discovery of high energy astrophysical neutrinos at 5.7 $ \sigma$ \cite{Aartsen:2014gkd}.  The 2 PeV event is the highest-energy neutrino ever observed.

High-energy neutrino telescopes are currently also providing data on neutrino oscillations measuring atmospheric neutrinos, commonly a background for astrophysical neutrino searches.   Using low energy samples, both ANTARES \cite{AdrianMartinez:2012ph} and IceCube/DeepCore \cite{Gross:2013iq} have measured the parameters $\theta_{23}$ and $ \Delta m^2_{23}$ in good agreement with existing data. PINGU, IceCube extension in the 10 GeV energy range, could measure the mass hierarchy and be sensitive to the Dirac phase \cite{Akhmedov:2012ah}. Such measurements exploit the occurrence of the matter effect on neutrinos through the Earth, both from the MSW and from the parametric resonance  \cite{Petcov:1998su}. Feasibility studies are currently ongoing both for PINGU and for ORCA \cite{Kouchner:2014epa} to establish if the energy and angular resolution required for the mass hierarchy search can be achieved. Neutrino telescopes are also sensitive to other fundamental properties such as Lorentz and CPT violation  \cite{Abbasi:2010kx}, or sterile neutrinos.

In conclusion, neutrinos from astrophysical sources have brought milestones in our knowledge of the weak interaction sector and of our Universe. The recent measurement of the pp neutrinos establishes the starting reaction chain of the main mechanism for energy production in the Sun. The theoretical progress in our understanding of neutrino propagation in supernovae brings essential knowledge for unravelling the r-process site(s) and for future supernova neutrino observations.
The measurement of the diffuse supernova neutrino background could be achieved with large scale detectors; while it remains challenging to observe cosmological neutrinos which would give an image 1 s after the Big-bang.
The recent  ultra-high energy neutrinos discovery opens a new window on the Universe and constitute an essential step to uncover cosmic rays origin. Since almost two decades neutrinos keep surprising us and will likely continue in the future.


\begin{thebibliography}{99}

 \bibitem{Pontecorvo:1957cp}% 
  B.~Pontecorvo,
  {\it Mesonium and anti-mesonium, 
   Sov.\ Phys.\ JETP }  \textbf{6}  (1957) 429 
  [{\it Zh.\ Eksp.\ Teor.\ Fiz.}  \textbf{33}  (1957) 549].
  %%CITATION = SPHJA,6,429;%%

%\cite{Fukuda:1998mi}
\bibitem{Fukuda:1998mi} 
  Y.~Fukuda {\it et al.},
 {\it Evidence for oscillation of atmospheric neutrinos,
  Phys.\ Rev.\ Lett. }  {\bf 81} (1998) 1562.
  %%CITATION = HEP-EX/9807003;%%

\bibitem{Maki:1962mu}% 
Z.~Maki, M.~Nakagawa,  and S.~Sakata, 
  {\it Remarks on the unified model of elementary particles, 
 Prog.\ Theor.\ Phys.} \textbf{28} (1962)  870.
  %%CITATION = PTPKA,28,870;%%   


%\cite{Wolfenstein:1977ue}
\bibitem{Wolfenstein:1977ue} 
  L.~Wolfenstein,
{\it Neutrino Oscillations in Matter,
  Phys.\ Rev.\ D} {\bf 17} (1978) 2369.
  %%CITATION = PHRVA,D17,2369;%%

%\cite{Mikheev:1986gs}
\bibitem{Mikheev:1986gs} 
  S.~P.~Mikheev and A.~Y.~Smirnov,
{\it Resonance Amplification of Oscillations in Matter and Spectroscopy of Solar Neutrinos,
  Sov.\ J.\ Nucl.\ Phys.}  {\bf 42} (1985) 913
   [Yad.\ Fiz.\  {\bf 42}, 1441 (1985)].
  %%CITATION = SJNCA,42,913;%%



%\cite{Volpe:2013kxa}
\bibitem{Volpe:2013kxa} 
  C.~Volpe,
{\it Open issues in neutrino astrophysics,
  Annalen Phys.} {\bf 525}, no. 8-9 (2013) 588 
  [arXiv:1303.1681].
  %%CITATION = ARXIV:1303.1681;%%

%\cite{Bethe:1939bt}
\bibitem{Bethe:1939bt} 
  H.~A.~Bethe,
{\it Energy production in stars,
  Phys.\ Rev.}  {\bf 55} (1939) 434.
  %%CITATION = PHRVA,55,434;%%

%\cite{Davis:1968cp}
\bibitem{Davis:1968cp} 
  R.~Davis, Jr., D.~S.~Harmer and K.~C.~Hoffman,
{\it Search for neutrinos from the sun,
  Phys.\ Rev.\ Lett.}  {\bf 20}  (1968) 1205.
  %%CITATION = PRLTA,20,1205;%%

%\cite{Bahcall:1968hc}
\bibitem{Bahcall:1968hc} 
  J.~N.~Bahcall, N.~A.~Bahcall and G.~Shaviv,
{\it Present status of the theoretical predictions for the Cl-36 solar neutrino experiment,
  Phys.\ Rev.\ Lett.}  {\bf 20} (1968)  1209.
  %%CITATION = PRLTA,20,1209;%%

%\cite{Robertson:2012ib}
\bibitem{Robertson:2012ib} 
  W.~C.~Haxton, R.~G.~Hamish Robertson and A.~M.~Serenelli,
{\it Solar Neutrinos: Status and Prospects,
  Ann.\ Rev.\ Astron.\ Astrophys.} {\bf 51} (2013) 21
  [arXiv:1208.5723].
  %%CITATION = ARXIV:1208.5723;%%






%\cite{Ahmad:2002jz}
\bibitem{Ahmad:2002jz} 
  Q.~R.~Ahmad {\it et al.},
{\it Direct evidence for neutrino flavor transformation from neutral current interactions in the Sudbury Neutrino Observatory,
  Phys.\ Rev.\ Lett.}  {\bf 89} (2002) 011301
  [nucl-ex/0204008].
  %%CITATION = NUCL-EX/0204008;%%


%\cite{Eguchi:2002dm}
\bibitem{Eguchi:2002dm} 
  K.~Eguchi {\it et al.},
{\it First results from KamLAND: Evidence for reactor anti-neutrino disappearance,
  Phys.\ Rev.\ Lett.}  {\bf 90}  (2003) 021802
  [hep-ex/0212021].
  %%CITATION = HEP-EX/0212021;%%

%\cite{Collaboration:2011nga}
\bibitem{Collaboration:2011nga} 
  G.~Bellini {\it et al.},
{\it First evidence of pep solar neutrinos by direct detection in Borexino,
  Phys.\ Rev.\ Lett.}  {\bf 108}  (2012) 051302
  [arXiv:1110.3230].
  %%CITATION = ARXIV:1110.3230;%%

%\cite{Arpesella:2008mt}
\bibitem{Arpesella:2008mt} 
  C.~Arpesella {\it et al.},
 {\it Direct Measurement of the Be-7 Solar Neutrino Flux with 192 Days of Borexino Data,
  Phys.\ Rev.\ Lett.}  {\bf 101}  (2008) 091302
  [arXiv:0805.3843].
  %%CITATION = ARXIV:0805.3843;%%

%\cite{Bellini:2014uqa}
\bibitem{Bellini:2014uqa} 
  G.~Bellini {\it et al.},
{\it Neutrinos from the primary proton-proton fusion process in the Sun,
  Nature} {\bf 512}, no. 7515 (2014)  383.
  %%CITATION = NATUA,512,383;%%



%\cite{Friedland:2004pp}
\bibitem{Friedland:2004pp} 
  A.~Friedland, C.~Lunardini and C.~Pena-Garay,
{\it Solar neutrinos as probes of neutrino matter interactions,
  Phys.\ Lett.\ B} {\bf 594} (2004) 347 
  [hep-ph/0402266].
  %%CITATION = HEP-PH/0402266;%%


  \bibitem{Suzuki:2008zzf}% 
A.~Suzuki, 
{\it The 20th anniversary of SN1987A,
 J.\ Phys.\ Conf.\ Ser.} \textbf{120}  (2008) 072001.
  %%CITATION = 00462,120,072001;%%

%\cite{Vissani:2014doa}
\bibitem{Vissani:2014doa} 
  F.~Vissani,
{\it Comparative analysis of SN1987A antineutrino fluence},
  arXiv:1409.4710.

\bibitem{Longo:1987ub} 
   M.~J.~Longo,
{\it Tests of relativity from  SN1987a,
  Phys.\ Rev.\ D} {\bf 36} (1987)  3276.
  %%CITATION = PHRVA,D36,3276;%%

%\cite{Autiero:2007zj}
\bibitem{Autiero:2007zj} 
  D.~Autiero {\it et al.},
{\it Large underground, liquid based detectors for astro-particle physics in Europe: Scientific case and prospects,
  JCAP} {\bf 0711} (2007) 011
  [arXiv:0705.0116].
  %%CITATION = ARXIV:0705.0116;%%

%\cite{Scholberg:2012id}
\bibitem{Scholberg:2012id} 
  K.~Scholberg,
{\it Supernova Neutrino Detection,
  Ann.\ Rev.\ Nucl.\ Part.\ Sci.}  {\bf 62} (2012) 81
  [arXiv:1205.6003].
  %%CITATION = ARXIV:1205.6003;%%


\bibitem{Beacom:2010kk}% 
 J.~F.~Beacom, 
{\it The Diffuse Supernova Neutrino Background,
  Ann.\ Rev.\ Nucl.\ Part.\ Sci.} \textbf{60}  (2010) 439.
  %%CITATION = ARXIV:1004.3311;%% 


 \bibitem{Lunardini:2010ab}% 
C.~Lunardini, 
{\it Diffuse supernova neutrinos at underground laboratories},
  arXiv:1007.3252.
  %%CITATION = ARXIV:1007.3252;%%


%\cite{Janka:2012wk}
\bibitem{Janka:2012wk} 
  H.~T.~Janka,
{\it Explosion Mechanisms of Core-Collapse Supernovae,
  Ann.\ Rev.\ Nucl.\ Part.\ Sci.}  {\bf 62} (2012) 407 
  [arXiv:1206.2503].
  %%CITATION = ARXIV:1206.2503;%%

%\cite{Burrows:2012ew}
\bibitem{Burrows:2012ew} 
  A.~Burrows,
{\it Colloquium: Perspectives on core-collapse supernova theory,
  Rev.\ Mod.\ Phys.}  {\bf 85} (2013) 245
  [arXiv:1210.4921].
  %%CITATION = ARXIV:1210.4921;%%


%\cite{Tamborra:2014aua}
\bibitem{Tamborra:2014aua} 
  I.~Tamborra {\it et al.},
  {\it Self-sustained asymmetry of lepton-number emission: A new phenomenon during the supernova shock-accretion phase in three dimensions,
  Astrophys.\ J.}  {\bf 792}  (2014) 96
  [arXiv:1402.5418].
  %%CITATION = ARXIV:1402.5418;%%


%\cite{Volpe:2014yqa}
\bibitem{Volpe:2014yqa} 
  See articles in the Focus Issue,
{\it Nucleosynthesis and the role of neutrinos: state of the art and open issues,
  J.\ Phys.\ G} {\bf 41}, 040301 (2014).
  %%CITATION = JPAGA,G41,040301;%%


%\cite{Dighe:1999bi}
\bibitem{Dighe:1999bi} 
  A.~S.~Dighe and A.~Y.~Smirnov,
{\it Identifying the neutrino mass spectrum from the neutrino burst from a supernova,
  Phys.\ Rev.\ D} {\bf 62} (2000) 033007
  [hep-ph/9907423].
  %%CITATION = HEP-PH/9907423;%%
  
%\cite{Duan:2010bg}
\bibitem{Duan:2010bg} 
  H.~Duan, G.~M.~Fuller and Y.~Z.~Qian,
  {\it Collective Neutrino Oscillations,
  Ann.\ Rev.\ Nucl.\ Part.\ Sci.}  {\bf 60} (2010) 569
  [arXiv:1001.2799].
  %%CITATION = ARXIV:1001.2799;%%

%\cite{Duan:2009cd}
\bibitem{Duan:2009cd} 
  H.~Duan and J.~P.~Kneller,
{\it Neutrino flavour transformation in supernovae,
  J.\ Phys.\ G} {\bf 36} (2009) 113201
  [arXiv:0904.0974].
  %%CITATION = ARXIV:0904.0974;%%


%\cite{Pantaleone:1992eq}
\bibitem{Pantaleone:1992eq} 
  J.~T.~Pantaleone,
 {\it Neutrino oscillations at high densities,
  Phys.\ Lett.\ B} {\bf 287} (1992) 128.
  %%CITATION = PHLTA,B287,128;%%









%\cite{Raffelt:2007xt}
\bibitem{Raffelt:2007xt} 
  G.~G.~Raffelt and A.~Y.~Smirnov,
  {\it Adiabaticity and spectral splits in collective neutrino transformations,
  Phys.\ Rev.\ D} {\bf 76} (2007) 125008
  [arXiv:0709.4641].
  %%CITATION = ARXIV:0709.4641;%%

%\cite{Galais:2011gh}
\bibitem{Galais:2011gh} 
  S.~Galais and C.~Volpe,
{\it The neutrino spectral split in core-collapse supernovae: a magnetic resonance phenomenon,
  Phys.\ Rev.\ D} {\bf 84} (2011) 085005
  [arXiv:1103.5302].
  %%CITATION = ARXIV:1103.5302;%%


%\cite{Akhmedov:2014ssa}
\bibitem{Akhmedov:2014ssa} 
  E.~Akhmedov, J.~Kopp and M.~Lindner,
{\it Decoherence by wave packet separation and collective neutrino oscillations},
  arXiv:1405.7275.
  %%CITATION = ARXIV:1405.7275;%%

%\cite{Balantekin:2006tg}
\bibitem{Balantekin:2006tg} 
  A.~B.~Balantekin and Y.~Pehlivan,
{\it Neutrino-Neutrino Interactions and Flavor Mixing in Dense Matter,
  J.\ Phys.\ G} {\bf 34}  (2007) 47
  [astro-ph/0607527].
  %%CITATION = ASTRO-PH/0607527;%%

%\cite{Volpe:2013uxl}
\bibitem{Volpe:2013uxl} 
  C.~Volpe, D.~Vaananen and C.~Espinoza,
{\it Extended evolution equations for neutrino propagation in astrophysical and cosmological environments,
  Phys.\ Rev.\ D} {\bf 87} (2013) 113010
  [arXiv:1302.2374].
  %%CITATION = ARXIV:1302.2374;%%






%\cite{Vlasenko:2013fja}
\bibitem{Vlasenko:2013fja} 
  A.~Vlasenko, G.~M.~Fuller and V.~Cirigliano,
{\it Neutrino Quantum Kinetics,
  Phys.\ Rev.\ D} {\bf 89}  (2014) 105004
  [arXiv:1309.2628].
  %%CITATION = ARXIV:1309.2628;%%

%\cite{Sigl:1992fn}
\bibitem{Sigl:1992fn} 
  G.~Sigl and G.~Raffelt,
{\it General kinetic description of relativistic mixed neutrinos,
  Nucl.\ Phys.\ B} {\bf 406}  (1993) 423.
  %%CITATION = NUPHA,B406,423;%% 

 
  %\cite{McKellar:1992ja}
\bibitem{McKellar:1992ja} 
  B.~H.~J.~McKellar and M.~J.~Thomson,
{\it Oscillating doublet neutrinos in the early universe,
  Phys.\ Rev.\ D} {\bf 49}  (1994) 2710.
  %%CITATION = PHRVA,D49,2710;%%




%\cite{Serreau:2014cfa}
\bibitem{Serreau:2014cfa} 
  J.~Serreau and C.~Volpe,
 {\it Neutrino-antineutrino correlations in dense anisotropic media},
  arXiv:1409.3591.

%\cite{Vlasenko:2014bva}
\bibitem{Vlasenko:2014bva} 
  A.~Vlasenko, G.~M.~Fuller and V.~Cirigliano,
{\it Prospects for Neutrino-Antineutrino Transformation in Astrophysical Environments},
  arXiv:1406.6724.
  %%CITATION = ARXIV:1406.6724;%%

%\cite{Pehlivan:2011hp}
\bibitem{Pehlivan:2011hp} 
  Y.~Pehlivan, A.~B.~Balantekin, T.~Kajino and T.~Yoshida,
{\it Invariants of Collective Neutrino Oscillations,
  Phys.\ Rev.\ D} {\bf 84} (2011) 065008
  [arXiv:1105.1182].
  %%CITATION = ARXIV:1105.1182;%%

%\cite{Bardeen:1957mv}
\bibitem{Bardeen:1957mv} 
  J.~Bardeen, L.~N.~Cooper and J.~R.~Schrieffer,
{\it Theory of superconductivity,
  Phys.\ Rev.}  {\bf 108}, 1175 (1957).
  %%CITATION = PHRVA,108,1175;%%

%\cite{Vaananen:2013qja}
\bibitem{Vaananen:2013qja} 
  D.~Vaananen and C.~Volpe,
{\it Linearizing neutrino evolution equations including neutrino-antineutrino pairing correlations,
  Phys.\ Rev.\ D} {\bf 88} (2013) 065003 
  [arXiv:1306.6372].
  %%CITATION = ARXIV:1306.6372;%%




%\cite{Abe:2011fz}
\bibitem{Abe:2011fz} 
  Y.~Abe {\it et al.},
  {\it Indication for the disappearance of reactor electron antineutrinos in the Double Chooz experiment,
  Phys.\ Rev.\ Lett.}  {\bf 108} (2012) 131801
  [arXiv:1112.6353].
  %%CITATION = ARXIV:1112.6353;%%

%\cite{An:2012eh}
\bibitem{An:2012eh} 
  F.~P.~An {\it et al.},
 {\it Observation of electron-antineutrino disappearance at Daya Bay,
  Phys.\ Rev.\ Lett.}  {\bf 108} (2012) 171803
   [arXiv:1203.1669].
  %%CITATION = ARXIV:1203.1669;%%

%\cite{Ahn:2012nd}
\bibitem{Ahn:2012nd} 
  J.~K.~Ahn {\it et al.},
{\it Observation of Reactor Electron Antineutrino Disappearance in the RENO Experiment,
  Phys.\ Rev.\ Lett.}  {\bf 108} (2012) 191802 
  [arXiv:1204.0626].
  %%CITATION = ARXIV:1204.0626;%%


%\cite{Serpico:2011ir}
\bibitem{Serpico:2011ir} 
  P.~D.~Serpico {\it et al.},
{\it Probing the neutrino mass hierarchy with the rise time of a supernova burst,
  Phys.\ Rev.\ D} {\bf 85} (2012) 085031
  [arXiv:1111.4483].
  %%CITATION = ARXIV:1111.4483;%%


%\cite{Gava:2009pj}
\bibitem{Gava:2009pj} 
  J.~Gava, J.~Kneller, C.~Volpe and G.~C.~McLaughlin,
{\it A Dynamical collective calculation of supernova neutrino signals,
  Phys.\ Rev.\ Lett.} {\bf 103}  (2009) 071101
  [arXiv:0902.0317].
  %%CITATION = ARXIV:0902.0317;%%


%\cite{deGouvea:2012hg}
\bibitem{deGouvea:2012hg} 
  A.~de Gouvea and S.~Shalgar,
{\it Effect of Transition Magnetic Moments on Collective Supernova Neutrino Oscillations,
  JCAP} {\bf 1210}  (2012) 027
  [arXiv:1207.0516].
  %%CITATION = ARXIV:1207.0516;%%



%\cite{Balantekin:2007es}
\bibitem{Balantekin:2007es} 
  A.~B.~Balantekin, J.~Gava and C.~Volpe,
{\it Possible CP-Violation effects in core-collapse Supernovae,
  Phys.\ Lett.\ B} {\bf 662}, 396 (2008)
  [arXiv:0710.3112].
  %%CITATION = ARXIV:0710.3112;%%

%\cite{Gava:2008rp}
\bibitem{Gava:2008rp} 
  J.~Gava and C.~Volpe,
{\it Collective neutrinos oscillation in matter and CP-violation,
  Phys.\ Rev.\ D} {\bf 78}, 083007 (2008)
  [arXiv:0807.3418].
  %%CITATION = ARXIV:0807.3418;%%

%\cite{Kneller:2009vd}
\bibitem{Kneller:2009vd} 
  J.~P.~Kneller and G.~C.~McLaughlin,
{\it Three Flavor Neutrino Oscillations in Matter: Flavor Diagonal Potentials, the Adiabatic Basis and the CP phase,
  Phys.\ Rev.\ D} {\bf 80}, 053002 (2009)
  [arXiv:0904.3823].
  %%CITATION = ARXIV:0904.3823;%%


%\cite{Pehlivan:2014zua}
\bibitem{Pehlivan:2014zua} 
  Y.~Pehlivan, A.~B.~Balantekin and T.~Kajino,
{\it Neutrino Magnetic Moment, CP Violation and Flavor Oscillations in Matter,
  Phys.\ Rev.\ D} {\bf 90}, 065011 (214)
  [arXiv:1406.5489].
  %%CITATION = ARXIV:1406.5489;%%

%\cite{Vaananen:2011bf}
\bibitem{Vaananen:2011bf} 
  D.~Vaananen and C.~Volpe,
{\it The neutrino signal at HALO: learning about the primary supernova neutrino fluxes and neutrino properties,
  JCAP} {\bf 1110} (2011) 019
  [arXiv:1105.6225].
  %%CITATION = ARXIV:1105.6225;%%



%\cite{Aartsen:2014gkd}
\bibitem{Aartsen:2014gkd} 
  M.~G.~Aartsen {\it et al.},
{\it Observation of High-Energy Astrophysical Neutrinos in Three Years of IceCube Data,
  Phys.\ Rev.\ Lett.} {\bf 113}  (2014) 101101
  [arXiv:1405.5303].
  %%CITATION = ARXIV:1405.5303;%%


%\cite{AdrianMartinez:2012ph}
\bibitem{AdrianMartinez:2012ph} 
  S.~Adrian-Martinez {\it et al.},
{\it Measurement of Atmospheric Neutrino Oscillations with the ANTARES Neutrino Telescope,
  Phys.\ Lett.\ B} {\bf 714} (2012) 224 
  [arXiv:1206.0645].
  %%CITATION = ARXIV:1206.0645;%%


%\cite{Gross:2013iq}
\bibitem{Gross:2013iq} 
  A.~Gross,
{\it Atmospheric Neutrino Oscillations in IceCube,
  Nucl.\ Phys.\ Proc.\ Suppl.}  {\bf 237}  (2013) 272.
  %%CITATION = ARXIV:1301.4339;%%


%\cite{Akhmedov:2012ah}
\bibitem{Akhmedov:2012ah} 
  E.~K.~Akhmedov, S.~Razzaque and A.~Y.~Smirnov,
{\it Mass hierarchy, 2-3 mixing and CP-phase with Huge Atmospheric Neutrino Detectors,
  JHEP} {\bf 1302}  (2013) 082
  [Erratum-ibid. {\bf 1307}  (2013)  026].
  %%CITATION = ARXIV:1205.7071;%%

   \bibitem{Petcov:1998su}% 
S.~T.~Petcov, 
 {\it Diffractive - like (or parametric resonance - like?) enhancement of the earth (day - night) effect for solar neutrinos crossing the earth core,
   Phys.\ Lett.\ B } \textbf{434}  (1998) 321.
  %%CITATION = HEP-PH/9805262;%% 

%\cite{Kouchner:2014epa}
\bibitem{Kouchner:2014epa} 
  A.~Kouchner,
{\it Next-generation atmospheric neutrino experiments,
  Phys.\ Dark Univ.} {\bf 4}, 60 (2014).






%\cite{Abbasi:2010kx}
\bibitem{Abbasi:2010kx} 
  R.~Abbasi {\it et al.},
 {\it Search for a Lorentz-violating sidereal signal with atmospheric neutrinos in IceCube,
  Phys.\ Rev.\ D} {\bf 82}  (2010) 112003
  [arXiv:1010.4096].
  %%CITATION = ARXIV:1010.4096;%%

\end{thebibliography}
\end{document}